\documentclass[twocolumn]{el-author}

\newcommand{\ua}{\uparrow}
\newcommand{\nc}{\newcommand}
\nc{\da}{\downarrow} \nc{\hc}{\hat{c}} \nc{\hS}{\hat{S}}
\nc{\bra}{\langle} \nc{\ket}{\rangle} \nc{\eq}{equation (\ref}
\nc{\h}{\hat} \nc{\hT}{\h{T}}\nc{\be}{\begin{eqnarray}}
\nc{\ee}{\end{eqnarray}}\nc{\rd}{\textrm{d}}\nc{\e}{eqnarray}\nc{\hR}{\hat{R}}\nc{\Tr}{\mathrm{Tr}}
\nc{\tS}{\tilde{S}}\nc{\tr}{\mathrm{tr}}\nc{\8}{\infty}\nc{\lgs}{\bra\ua,\phi|}\nc{\rgs}{|\ua,\phi\ket}
\nc{\hU}{\hat{U}}\nc{\lfs}{\bra\phi|}\nc{\rfs}{|\phi\ket}\nc{\hZ}{\hat{Z}}\nc{\hd}{\hat{d}}\nc{\mD}{\mathcal{D}}
\nc{\bd}{\bar{d}}\nc{\bc}{\bar{c}}\nc{\mc}{\mathcal}\nc{\ea}{eqnarray}\nc{\mG}{\mathcal{G}}\nc{\bce}{\begin{center}}
\nc{\ece}{\end{center}}
\date{30th November 2017}

\begin{document}

\title{High-{\it Q} nanocavities in semiconductor-based three-dimensional photonic crystals}

\author{S. Takahashi, T. Tajiri, K. Watanabe, Y. Ota, S. Iwamoto, and Y. Arakawa}

\abstract{
We experimentally demonstrated high quality factors ({\it Q}-factors) of nanocavities in three-dimensional photonic crystals by increasing the in-plane area of the structure.
Entire structures made of GaAs were fabricated by a micro-manipulation technique, and the nanocavities contained InAs self-assembled quantum dots that emitted near-infrared light.
The obtained {\it Q}-factor was improved to 93,000, which is 2.4-times larger than that in a previous report of a three-dimensional photonic crystal nanocavity.
Due to this large {\it Q}-factor, we successfully observed a lasing oscillation from this cavity mode.
}

\maketitle


\section{Introduction}

Three-dimensional (3D) photonic crystals (PhCs) composed of periodically arranged materials having different refractive indices can exhibit complete photonic bandgaps (cPBGs), which prevent omni-directional propagation of light having any polarizations \cite{Yablonovitch,John,Lin,Vos,Yamamoto,Ishizaki}.
Point defects in 3D PhCs create localized states of light in the cPBG and work as nanocavities in which high quality factors ({\it Q}-factors) and small mode volumes are achievable, and this is expected to allow exploration of light-matter interactions or to realize basic elements required in 3D optical circuits.
The highest reported {\it Q}-factor among 3D PhC nanocavities was observed in a semiconductor-based woodpile structure fabricated using a micro-manipulation technique \cite{Aniwat}.
However, the observed {\it Q}-factor was only $\sim$ 38,500 due to the limited area of the structure in the in-plane directions.
In this study, we succeeded in fabricating 3D woodpile structures having large in-plane areas that were 4-times larger than that of the previous 3D PhC \cite{Aniwat}.
We evaluated the {\it Q}-factors at the transparency pump power \cite{Aniwat}, and observed that the {\it Q}-factor was as high as 93,000, which is 2.4-times larger than that in the previous report.
Due to this high {\it Q}-factor, we observed a lasing oscillation from this cavity mode.
Such high {\it Q}-factors in 3D PhC nanocavities are expected to lead to thresholdless lasing oscillation for 3D confined light.


\section{Sample fabrication}

Using molecular beam epitaxy, after growing a GaAs buffer layer on a GaAs substrate, we grew a sacrificial Al$_{0.7}$Ga$_{0.3}$As layer having a thickness of 1.5 $\mu$m, followed by a GaAs slab layer having a thickness of 150 nm.
In-plane patterns having a 11 $\mu$m $\times$ 11 $\mu$m area, which is 4-times larger than area of a previously reported structure \cite{Aniwat}, were drawn on the GaAs slab by electron beam lithography.
In the pattern, rods having a width of 130 nm were aligned at a period of 500 nm.
The patterns were then transferred to the GaAs slab by inductively coupled plasma ion etching.
Finally, the sacrificial layer was removed by a wet etching process using a hydrofluoric acid solution to form suspended structures, as shown in Fig. 1(a).
Note that when the length of the rods was longer than 11 $\mu$m, the neighboring rods touched each other after the wet etching process due to the surface tension of the etchant.
In addition to these passive plates, we also grew another GaAs slab layer containing three-layer-stacked InAs self-assembled quantum dots (QDs) on another GaAs substrate.
The QDs, having a diameter of $\sim$ 30 nm, a height of $\sim$ 5 nm, and a density of $\sim$ 1.0$\times$10$^{10}$ /cm$^{2}$, were used as near-infrared light emitters.
Figure 1(b) shows a typical photoluminescence (PL) spectrum from the bare QDs without any structure excited by continuous wave (CW) laser light having a wavelength of 980 nm and a power of 196 $\mu$W.
This wafer was processed in the same way as the processes used for the passive plates to fabricate suspended active plates.
As shown in Fig. 1(c), a point defect having an area of 1.13 $\mu$m $\times$ 1.13 $\mu$m was located at the center of the active plates.


These suspended plates were picked up and stacked one-by-one using a micro-manipulation technique \cite{Aoki1,Aoki2} to form a woodpile structure in which four different plates formed a single period.
As guides for stacking, we prepared three vertical posts having a height of 7 $\mu$m by anisotropically etching a GaAs substrate.
We fabricated four 3D PhC nanocavities by stacking firstly 16 lower passive plates, then an active plate, and finally different numbers of upper passive plates, namely, 4, 8, 12, and 16.
The fabricated 3D PhC having 16 upper plates is shown in Fig. 1(d).
Note that the guiding posts were tapered about 150 nm from the top to the bottom non-linearly, and the stacking accuracy in the in-plane directions between the neighboring plates was $<$ 50 nm for the bottom region and $<$ 100 nm for the top region.
In the stacking direction, the accuracy was $<$ 10 nm.


\begin{figure}
\centering{\includegraphics[width=85mm]{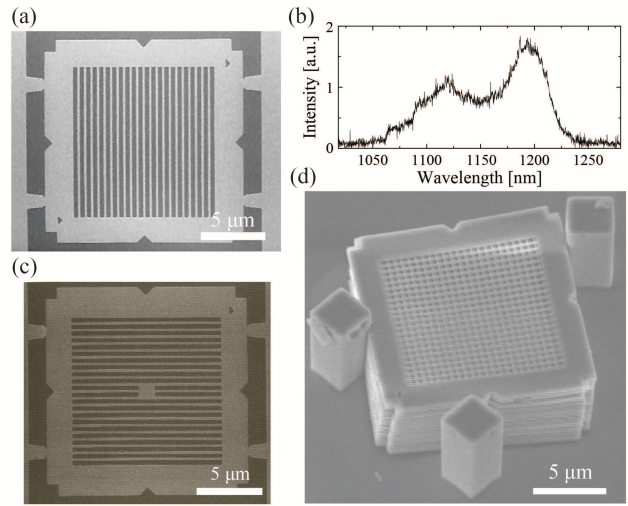}}
\caption{
(a) Scanning Electron Microscope (SEM) image for one of the four kinds of passive plates constructing a woodpile structure. 
(b) Typical PL spectrum from bare QDs without any structure. 
(c) SEM image of a typical active plate containing QDs and a point defect. 
(d) Stacked sample having 16 lower plates, a single active plate, and 16 upper plates. 
The three posts are guides for stacking.
The micro-manipulation was performed under the SEM observation, with the samples tilted by 45 degrees to the incident electron beam.
The bright part of the 3D PhC was caused by the shadow of the guide post for the electron beam.
\source{}}
\end{figure}


\section{Experimental setup}

PL measurements were performed at 20 K using a pulsed semiconductor laser having a repetition rate of 25 kHz to excite the wetting layer beneath the QDs in the nanocavities.
Note that the wavelength of the laser light was 905 nm, which was outside the cPBG of the 3D PhC, and the pulse duration of 8 ns was longer than the exciton life time of $<$ 1 ns, so that the excitation laser pulse was considered to be quasi-CW.
Some of the excited carriers relaxed into a discrete energy state in the QDs and recombined to emit near-infrared light.
The laser light was focused on the PhC by a 50$\times$ objective lens having a numerical aperture of 0.65, and the PL signals were resolved in frequency space by a monochromator and detected by an InGaAs photodiode array.
The instrument response function of the system was measured to be 0.0144 nm by using a CW laser having a wavelength of 1270 nm.


\section{{\it Q}-factor evaluation}

As assumed in the previous report \cite{Aniwat}, a {\it Q}-factor can be evaluated at a transparency pump power that is just below the threshold power.
We adopted this method for evaluating the {\it Q}-factors in order to compare them with the previous report.
The PL measurement showed a cavity mode around wavelength of 1231 nm for a 3D PhC nanocavity having 12 upper layers.
Figure 2(a) shows the PL spectrum at an input power of 4.35 $\mu$W, which is just below the threshold power of 4.59 $\mu$W discussed in the next section.
To evaluate the {\it Q}-factor accurately, we fitted the peak using a Voigt function, which is a convolution of a Lorentzian and a Gaussian function.
Note that the linewidth of the Gaussian function was fixed at 0.0144 nm due to the instrument response function.
We finally obtained a Lorentzian function linewidth of 0.0133 nm, corresponding to {\it Q} $\sim$ 93,000.
This value of the {\it Q}-factor is 2.4-times larger than that in the previous report \cite{Aniwat}.


The evaluated {\it Q}-factors are plotted as black open squares in Fig. 2(b) as a function of the number of upper plates.
Note that the {\it Q} $\sim$ 1,500 in the PhC having 4 upper plates was only evaluated by fitting a simple Lorentzian function, because the nanocavity did not show any lasing oscillation due to the weak confinement of light in the stacking direction.
We observed lasing oscillations in the PhCs having 8 and 16 upper plates, and the {\it Q}-factors were 89,000 and 62,000, respectively.
The {\it Q}-factors numerically calculated using a finite-difference time-domain method are also plotted in Fig. 2(b).
Among several cavity modes in the calculation, we chose cavity modes showing the largest {\it Q}-factors in the PhCs having 4 and 8 upper plates.
These calculated results agree well with the measured {\it Q}-factors as long as the number of upper plates is less than 8.
When the number of upper plates $>$ 8, the measured {\it Q}-factors were saturated at {\it Q} $\sim$ 100,000.
This is probably because the cavity energy leaks due to fabrication imperfections in the 3D PhCs, particularly the in-plane stacking errors of 50-100 nm between the neighboring plates.
The fluctuations of the {\it Q}-factors in the PhCs having $>$ 8 upper plates were also due to the stacking errors, which depended on the individually fabricated PhCs.
The relatively small stacking error in the PhC having 12 upper plates resulted in the largest {\it Q}-factor.
To achieve further improvement of the {\it Q}-factor, new stacking methods with high accuracy will be required.
For example, instead of the tapered vertical posts used as stacking guides, precise guides in the in-plane directions can be fabricated by electron-beam lithography, allowing plates to be stacked not in the vertical direction but in the in-plane direction \cite{Tajiri}.


\begin{figure}
\centering{\includegraphics[width=85mm]{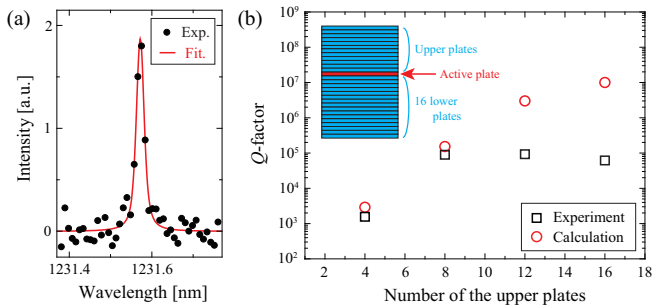}}
\caption{
(a) PL spectrum around wavelength of 1231 nm for an input power of 4.35 $\mu$W, which is just below the threshold power of 4.59 $\mu$W.
The {\it Q}-factor was evaluated by fitting a Voigt function, shown as a red curve.
(b) Experimentally and numerically obtained {\it Q}-factors as a function of the number of upper plates.
The measured {\it Q}-factors saturated due to the stacking errors.
The evaluation error for each experimental plot was smaller than the size of the symbol.
Inset shows a schematic diagram of the layered 3D PhC composed of 16 passive plates, an active plate, and various numbers of upper plates.
\source{}}
\end{figure}


\section{Lasing oscillation}
For the cavity mode shown in Fig. 2(a), the integrated intensity as a function of the averaged input power is shown on a linear scale and a logarithmic scale in Fig. 3(a) and (b), respectively.
The clear non-linear increase in Fig. 3(a) and the characteristic S shape of the intensity in Fig. 3(b) suggest a lasing oscillation in this cavity mode.
From a conventional linear fitting in Fig. 3(a), the threshold power was determined to be 4.59 $\mu$W.
The fitted blue curve for the integrated intensity in Fig. 3(b) is based on rate equations without non-radiative recombination processes \cite{Strauf}, resulting in the spontaneous coupling factor $\beta$ = 0.07.
In Fig. 3(b), we also plotted the linewidth obtained by simple Lorentzian fittings as a function of the averaged input power.
The linewidth became narrower with higher input power.
These results showing the non-linear increase and the characteristic evolution of the intensity, as well as the narrowing of the linewidth, indicate the lasing oscillation of the mode \cite{Strauf}.


\begin{figure}
\centering{\includegraphics[width=85mm]{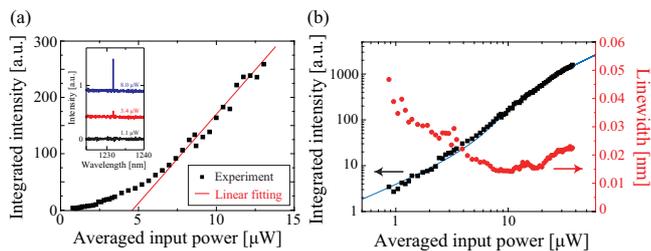}}
\caption{
(a) Linear plot of the integrated intensity as a function of the averaged input power for the peak at a wavelength of 1231 nm.
Linear fitting in red shows a threshold power of 4.59 $\mu$W.
Inset shows PL spectra for various averaged input powers.
(b) Logarithmic plot of the integrated intensity (black) and linear plot of the linewidth (red) for the peak at a wavelength of 1231 nm. 
The blue curve is fitted using rate equations.
Note that the linewidth of the instrument response function was 0.0144 nm.
\source{}}
\end{figure}


\section{Conclusion}

We experimentally improved the {\it Q}-factors of the nanocavities in the semiconductor-based 3D PhCs by increasing the in-plane area of the structures.
The {\it Q}-factors evaluated at the transparency pump power was as high as $\sim$ 93,000 in the near-infrared band.
This value is 2.4-times larger than the previous report of a 3D PhC nanocavity.
Due to this large {\it Q}-factor, a lasing oscillation from this cavity mode was observed.
We also found that the stacking errors in the current micro-manipulation technique prevent further improvement of the {\it Q}-factor.
Novel stacking methods with high accuracy are expected to allow the realization of more elaborate 3D PhC nanocavities, which will lead to thresholdless lasers or optical circuits including photon manipulation based on light-matter interactions.


\vskip3pt
\ack{This work was supported by a JSPS KAKENHI Grant-in-Aid for Scientific Research (No. JP16H06085), a Grant-in-Aid for Specially Promoted Research (No. JP15H05700), and a Grant-in-Aid for Scientific Research on Innovative Areas (No. JP15H05868).}

\vskip5pt
\noindent S. Takahashi, K. Watanabe, Y. Ota, S. Iwamoto, and Y. Arakawa (\textit{Institute for Nano Quantum Information Electronics, University of Tokyo, Tokyo, Japan})

\noindent S. Takahashi (\textit{Kyoto Institute of Technology, Kyoto, Japan})

\noindent T. Tajiri, S. Iwamoto, and Y. Arakawa (\textit{Institute of Industrial Science, University of Tokyo, Tokyo, Japan})

\noindent S. Takahashi and T. Tajiri contributed equally to this work.

\vskip3pt
\noindent E-mail: shuntaka@kit.ac.jp

\end{document}